\documentclass[12pt,preprint]{aastex631}

\usepackage{epsfig}

\begin{document}
\title{The Longest Delay: a 14.5 Yr Campaign to Determine the Third Time Delay in the Lensing Cluster SDSS~J1004+4112}
\author[0000-0001-9833-2959]{J.A.~Mu\~noz}
\affiliation{Departamento de Astronom\'{\i}a y Astrof\'{\i}sica, Universidad de Valencia, E-46100 Burjassot, Valencia, Spain}
\affiliation{Observatorio Astron\'omico, Universidad de Valencia, E-46980 Paterna, Valencia, Spain}

\author[0000-0001-6017-2961]{C.S.~Kochanek}
\affiliation{Department of Astronomy, The Ohio State University, 140 West 18th Avenue, Columbus OH 43210}
\affiliation{Center for Cosmology and AstroParticle Physics, The Ohio State University, 191 W. Woodruff Avenue, Columbus OH 43210 }
    
\author{J.~Fohlmeister}
\affiliation{Leibniz-Institut f\"ur Astrophysik Potsdam (AIP), An der Sternwarte 16, 14482 Potsdam, Germany}
\affiliation{Zentrum f\"ur Astronomie der Universit\"at Heidelberg, Astronomisches Rechen-Institut, M\"onchhofstr. 12-14, D-69120 Heidelberg, Germany}

\author[0000-0002-8365-7619]{J.~Wambsganss}
\affiliation{Zentrum f\"ur Astronomie der Universit\"at Heidelberg, Astronomisches Rechen-Institut, M\"onchhofstr. 12-14, D-69120 Heidelberg, Germany}

\author[0000-0002-7061-6519]{E.~Falco}
\affiliation{Harvard-Smithsonian Center for Astrophysics, 60 Garden St., Cambridge, MA, 02138, USA}

\author[0000-0002-6482-2180]{R.~For\'es-Toribio}
\affiliation{Departamento de Astronom\'{\i}a y Astrof\'{\i}sica, Universidad de Valencia, E-46100 Burjassot, Valencia, Spain}
\affiliation{Observatorio Astron\'omico, Universidad de Valencia, E-46980 Paterna, Valencia, Spain}

\begin{abstract}
We present new light curves for the four bright images of the five image
cluster-lensed quasar gravitational lens
system SDSS~J1004+4112. The light curves span 14.5 yr and
allow measurement of the time delay between the trailing bright quasar image D and
the leading image C.  When we fit all four light curves
simultaneously and combine the models using the Bayes information criterion, we find a time delay of
$\Delta t_{DC}= 2458.47 \pm 1.02$ days (6.73 yr), the longest ever measured for a gravitational lens.
For the other two independent time delays we obtain  
$\Delta t_{BC}=782.20 \pm 0.43$ days (2.14 yr) and $\Delta t_{AC}= 825.23 \pm 0.46$ days (2.26 yr),
in agreement with previous results.  The information criterion is needed
to weight the results for light curve models with different
polynomial orders for the intrinsic variability and
the effects of differential microlensing.  The results using the Akaike
information criterion are slightly different, but, in practice,
the absolute delay errors are all dominated by the $\sim 4\%$ cosmic variance in the delays
rather than the statistical or systematic measurement uncertainties. Despite the lens being a cluster, the quasar images show slow differential variability due to microlensing at the level of a few tenths of a magnitude.
\end{abstract}


\section{Introduction}\label{sec1}

SDSS J1004+4112 is a galaxy cluster lens at $z_l=0.68$ with four images forming a typical quad lens \citep{2003Natur.426..810I} and a faint central image \citep{2005PASJ...57L...7I,2008PASJ...60L..27I} of a single background quasar at $z_s=1.73$. The maximum image separation is 14.6~arcsec.
The system also has seven multiply imaged background galaxies at three different redshifts ($z=1.73$ , $2.74$ and $3.33$, \citealt{2005PASJ...57L...7I,2005ApJ...629L..73S,2009MNRAS.397..341L,2010PASJ...62.1017O}). There are also radio, infrared, ultraviolet and X-ray observations which have been used to study the wavelength-dependent quasar flux ratios, the cluster and background lensed galaxies and the mass of the cluster \citep{2006ApJ...647..215O,2009ApJ...702..472R,2011ApJ...739L..28J,2021MNRAS.505L..36M}.

The large image separations also lead to large time delays between the images. A monitoring campaign from 2003 December to 2006 June by \citet{2007ApJ...662...62F} led to the measurement of the time delay between images A and B, the brightest and second brightest images of the quasar, respectively. This delay
of $38.4\pm 2.0$~days is relatively short because images A and B are close to merging on a critical line with a separation of only 3\farcs8. An extended campaign from 2006 October to 2007 June \citep{2008ApJ...676..761F} allowed the determination of the delay between the image A and the leading image C ($\Delta t_{AC}=821.6\pm2.1$ days) and refined the time delay between A and B ($\Delta t_{AB}=40.6\pm1.8$ days). \citet{2008ApJ...676..761F}  also set a lower limit of $\Delta t_{DA}> 1250$~days on the delay of the fourth brightest image D.

An interesting feature of this slow release of delays was the insight it provides into the accuracy of cluster lens models.  Prior to the first delay measurement, \citet{2004ApJ...605...78O}
predicted AB delays from roughly $-8$ to $+26$~days and CD delays from $-1000$ to $+3000$~days.  \citet{2004AJ....128.2631W} predicted AB delays of roughly $0$ to $25$~days, CD delays of $800$ to $1700$ days, BD delays of $250$ to $950$~days and AD delays of $450$ to $950$~days.  Finally, \citet{2006PASJ...58..271K} predicted AB delays of roughly $4$ to $54$~days, CB delays of $0$ to $2000$~days, and CD delays of $200$ to $4600$~days.  Despite the tremendous range of these predictions, they did  not encompass the first delay measurement (a single outlier in \citet{2006PASJ...58..271K} was longer than the \citet{2007ApJ...662...62F} measurement, but there were no models consistent with it).
\citet{2007ApJ...662...62F} argued that the problem with these initial models was that they largely ignored the perturbing effects of galaxies on the lens structure and the delays.   In their models including galaxies, they found BC delays of $450$ to $1000$~days which were consistent with both a (wrong) initial estimate of the BC delay and the subsequent measurement of it in \cite{2008ApJ...676..761F}.  There have been three subsequent models including both time delay measurements: \citet{2009MNRAS.397..341L} predicted a CD delay of $\sim 1300$~days;
\citet{2010PASJ...62.1017O} predicted an AD delay of 1200 to 1350~days; and 
\citet{2015PASJ...67...21M} predicted CD delays of 1500 to 2700~days.

We are now able to measure this
last independent time delay after monitoring the system for 14.5 years, and we reveal below whether the three more recent predictions were more successful than the older attempts.
In Section \ref{sec2} we describe the observations and the extraction of the
light curves. The time delay measurements for the four bright images are presented in Section \ref{sec3}. We discuss the results in Section \ref{sec4}.

\section{Observations and Data Analysis}\label{sec2}

The new observations were all acquired at the Fred Lawrence Whipple Observatory (FLWO) 1.2m, on Mount Hopkins.  Between 2004 and 2010, the scheduled observer was provided with a standard observing script and asked to observe SDSS~J1004$+$4112 each night. Starting in July 2010, all observations switched to a robotic system and they were
automatically scheduled and executed.
Our standard observations consisted of two consecutive, unguided
300-sec exposures with the Sloan r filter using the Keplercam CCD.
It was binned by two, leading to $0\farcs76$ binned pixels. The
1.2m tracks extremely well without guiding, so the observations
were unguided. In the years after its installation in
1990, the surface of the 1.2m primary mirror visibly deteriorated leading
to a steady degradation in the delivered image quality to a median
full-width at half-maximum (FWHM) of 2.5~arcsec.
FLWO replaced the 1.2m primary with a new mirror in September 2012 which
improved the typical FWHM to 1.5~arcsec.
Observations were obtained only when the moon was at least 90 degrees
away from the target, clouds were absent, and the observed seeing
was (believed to be) better than 3 arcsec. This was to try and ensure
that the quasars were always detectable.

\begin{figure}
\begin{center}
\includegraphics[width=\textwidth]{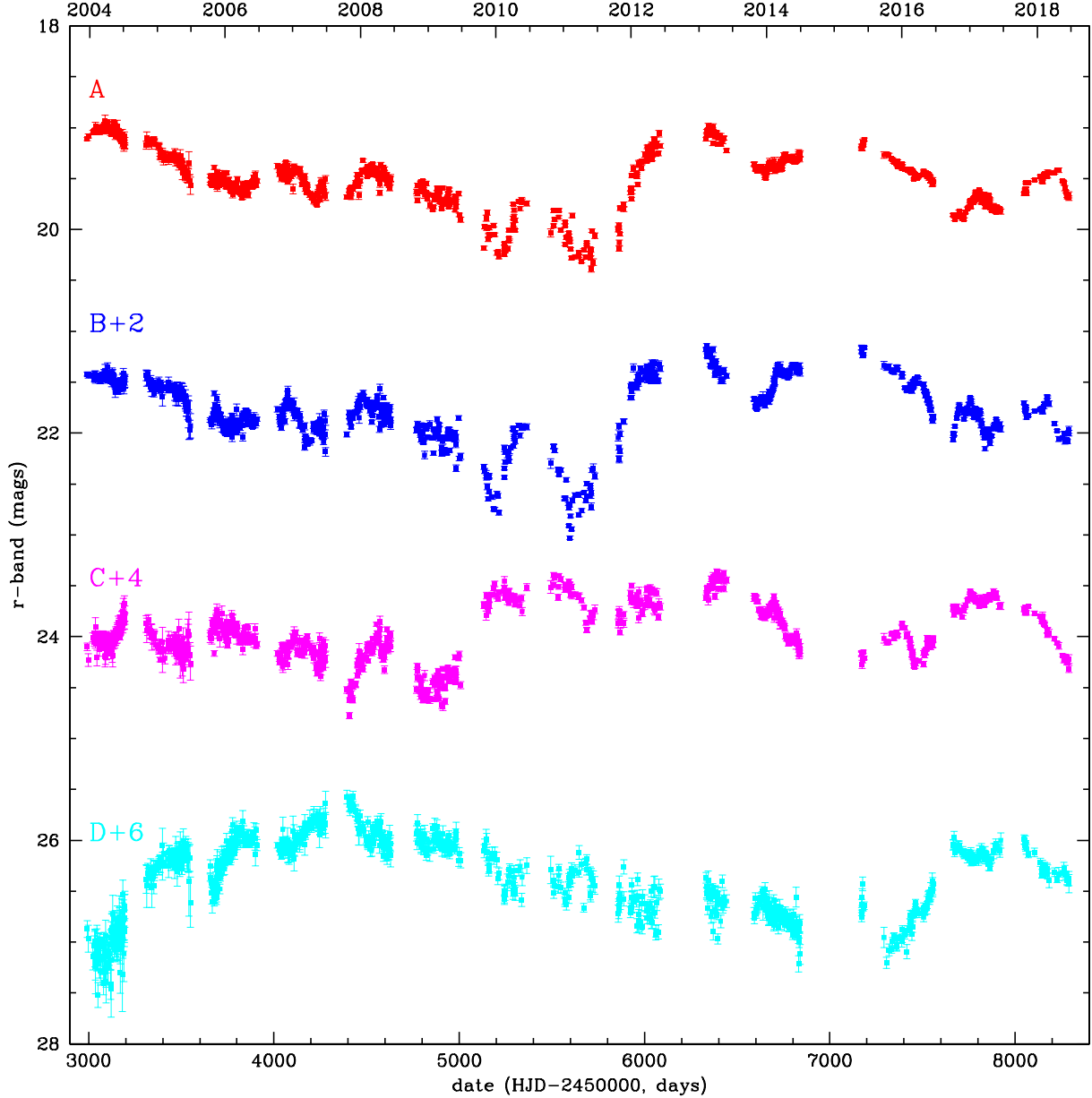}
\caption{\label{lc} The light curves of the four bright quasar images in
   SDSS~J1004$+$4112.  The B, C and D images are offset by 2, 4 and 6 magnitudes,
  respectively, for display purposes. The data span 14.5 years, from December 20, 2003 until June 21, 2018.}
\end{center}
\end{figure}

We analyzed all of the Keplercam observations, a total of 797 epochs, including the 
88 and 85 epochs already published in \citet{2007ApJ...662...62F} and \citet{2008ApJ...676..761F},
respectively.  For the new analysis we used difference imaging (\citealt{Alard1998},
\citealt{Alard2000}) light curves with the
 r-band reference image calibrated 
using the same 5 stars as in  \citet{2007ApJ...662...62F} and \citet{2008ApJ...676..761F}.
For the 173 reprocessed epochs we computed the differences between the new
and old values, finding a mean and variance for each
quasar image of $\Delta m_A = 0.096 \pm 0.040$, $ \Delta m_B = 0.058 \pm
0.058$, $\Delta m_C = 0.0047 \pm 0.062$, and $\Delta m_D= -0.10 \pm
0.07$~mag.
Table 1 gives the final light curves with 1018 epochs after removing epochs where
the seeing FWHM was worse than 5\arcsec\ or there were problems seen in a visual 
inspection of each image.
Fig.~1 shows the final light curves.

\section{Results}\label{sec3}\

We measure the time delays using the same basic procedures as in \citet{2007ApJ...662...62F} and \citet{2008ApJ...676..761F}. We
model the quasar light curve by a high order polynomial for the leading
image C combined with three additional polynomials for the differences
in microlensing between this reference image and each of the other
three images.  The assumption is that microlensing variability
generally occurs on longer time scales (see \citealt{2011ApJ...738...96M} for
the expectations for microlensing time scales) than the intrinsic variability
of the quasar, so the microlensing polynomials are lower order
than the quasar polynomials.  Given a choice of polynomial orders,
we can then compute a $\chi^2$ goodness of fit of the model to the
light curves as a function of the delays.  After initial pair wise
fits to pin down the delay ranges that needed to be explored, the
final fits were to all four images simultaneously, although we do
report the results for the pair wise fits.  Because we have
four images and two long delays, the seasonal gaps are all filled.
This avoids a common problem in time delay measurements where the
goodness of fit can be improved by using the seasonal gaps to reduce
the time period where the light curves overlap. For the final Joint fits we 
dropped the parts of the C ($JD>2457600$) and D ($JD< 2454600$) light curves which
will never overlap with the other images. After trimming
these data, we are left with $N=3473$ magnitudes to be fit.

To be specific, we model the quasar with a polynomial of order
$n=100 i + 50$ with $i=1$ to $5$ and the microlensing as polynomials
of order $m=1$ to $15$.  This leads to a family of 75 models. Clearly,
picking any particular model would be somewhat arbitrary, so we 
instead combine all the models using Bayesian methods.  Since the
models have very different numbers of parameters, we require an
information criterion for how models are penalized given 
their number of parameters $p=n+3m$.  We use the Akaike information
criterion (AIC), where the penalty added to the log-likelihood
($\ln L = -\chi^2/2$) is $p$, and the Bayes information criterion
where the penalty is $(p/2)\ln N$ where $N$ is the number of data.
The Bayes information criterion (BIC) penalizes new parameters much
more heavily since, for $N=3473$,
the BIC factor of $(p/2)\ln N = 4.08p$ is four times larger than the
AIC factor of just $p$.  Ideally
we will find similar results for both even though they will 
weight the 75 models very differently.  For the pair wise fits,
the procedures are the same but there is only one microlensing
polynomial instead of three.  For the AIC models, the likelihood
steadily increases as we increase the polynomial order,
while for the BIC models, the maximum likelihood model
has $n=350$ and $m=15$. This model has a $\chi^2=16839$ for
3078 degrees of freedom. 
The large $\chi^2$ is driven by outliers in the photometric data. It could be reduced 
eliminating outliers completely or by broadening their uncertainties, 
but that process always has a degree of arbitrariness. 
The effect of uniformly broadening the uncertainties to make the $\chi^2$ per degree of freedom unity 
would simply be to broaden the statistical uncertainties by $\sqrt{16839/3078}\sim2.3$, 
which would still be less than the dominant uncertainties created by cosmic variance as we explain below.

Table~2 presents the results for all six image combinations, both information
criteria (AIC and BIC) and either fitting all 4 light curves simultaneously (Joint) or 
doing each pair individually (Pair). For the Joint fits, all 6 delay distributions
can be directly calculated from the distributions for the three lags actually used
as parameters in the fit.  The labeling of the delays is that image $i$
lags image $j$ by $\Delta t_{ij}$ where the overall ordering is that C varies first, 
followed by B, then A and finally D.  Figure 2 shows the AIC and BIC probability 
distributions for the joint fits to the BC, AC and DC delays

\begin{figure}
\centering
\begin{tabular}{ccc}
\includegraphics[width=5.75cm]{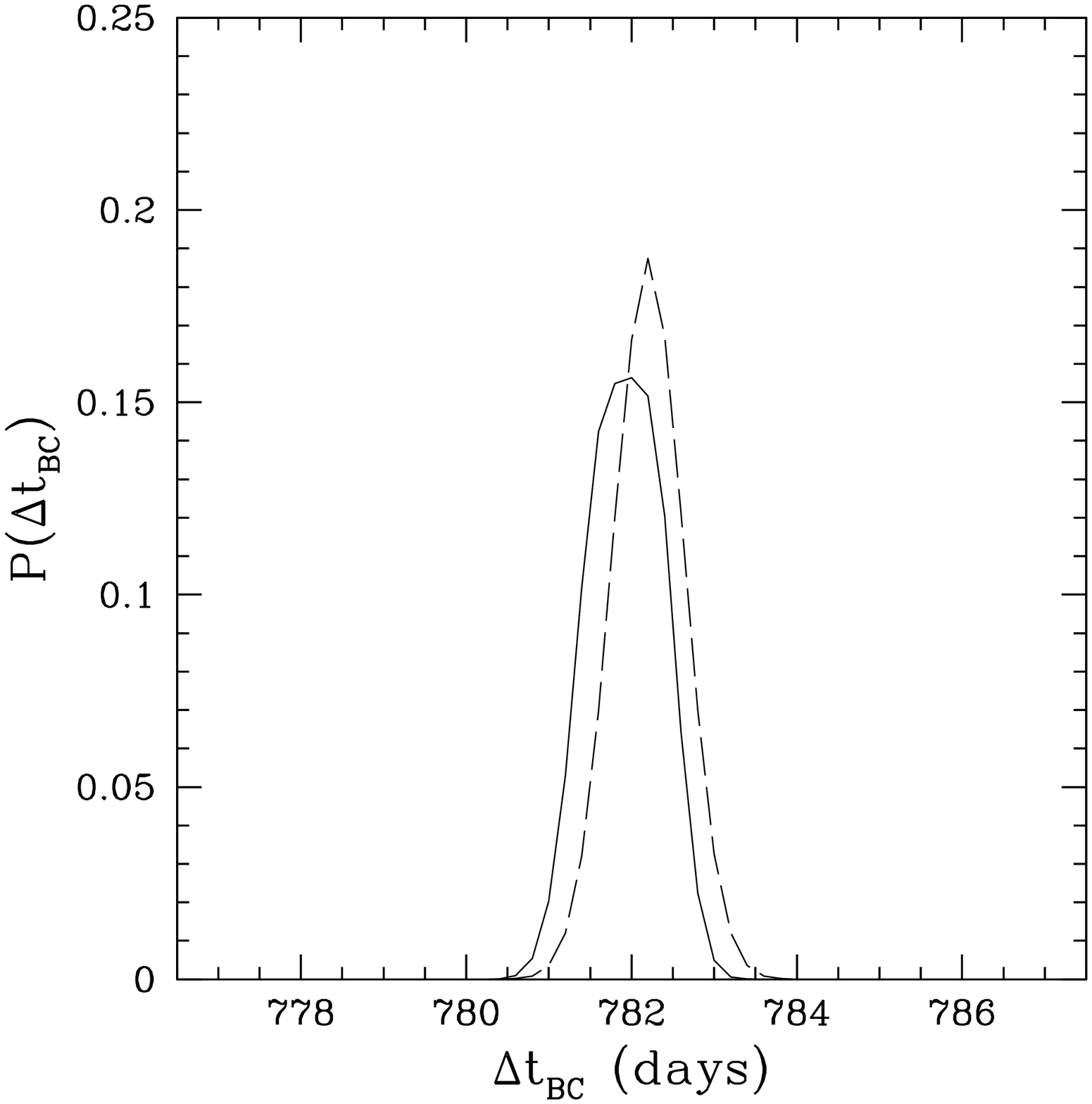} &
\includegraphics[width=5.75cm]{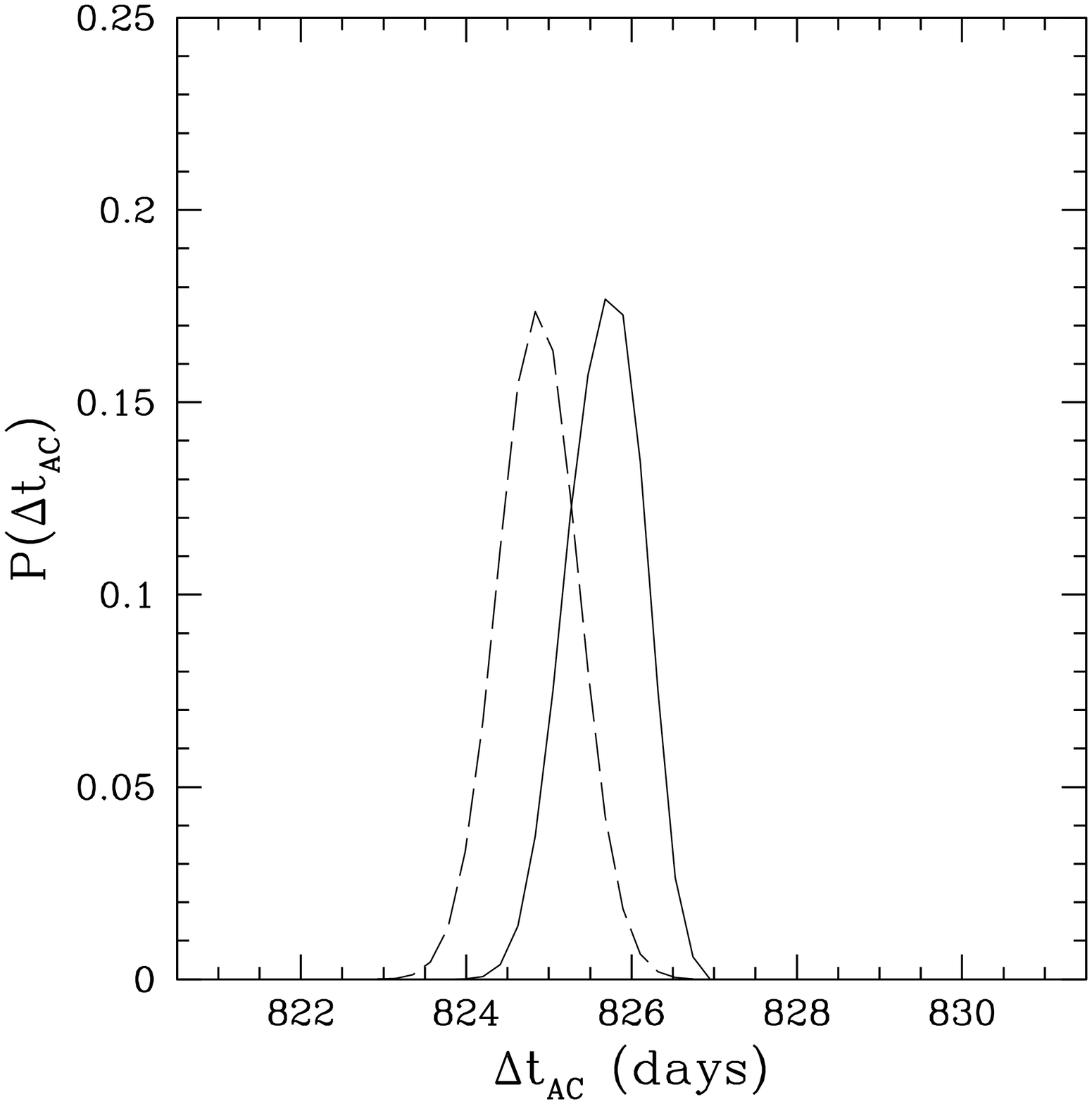} &
\includegraphics[width=5.75cm]{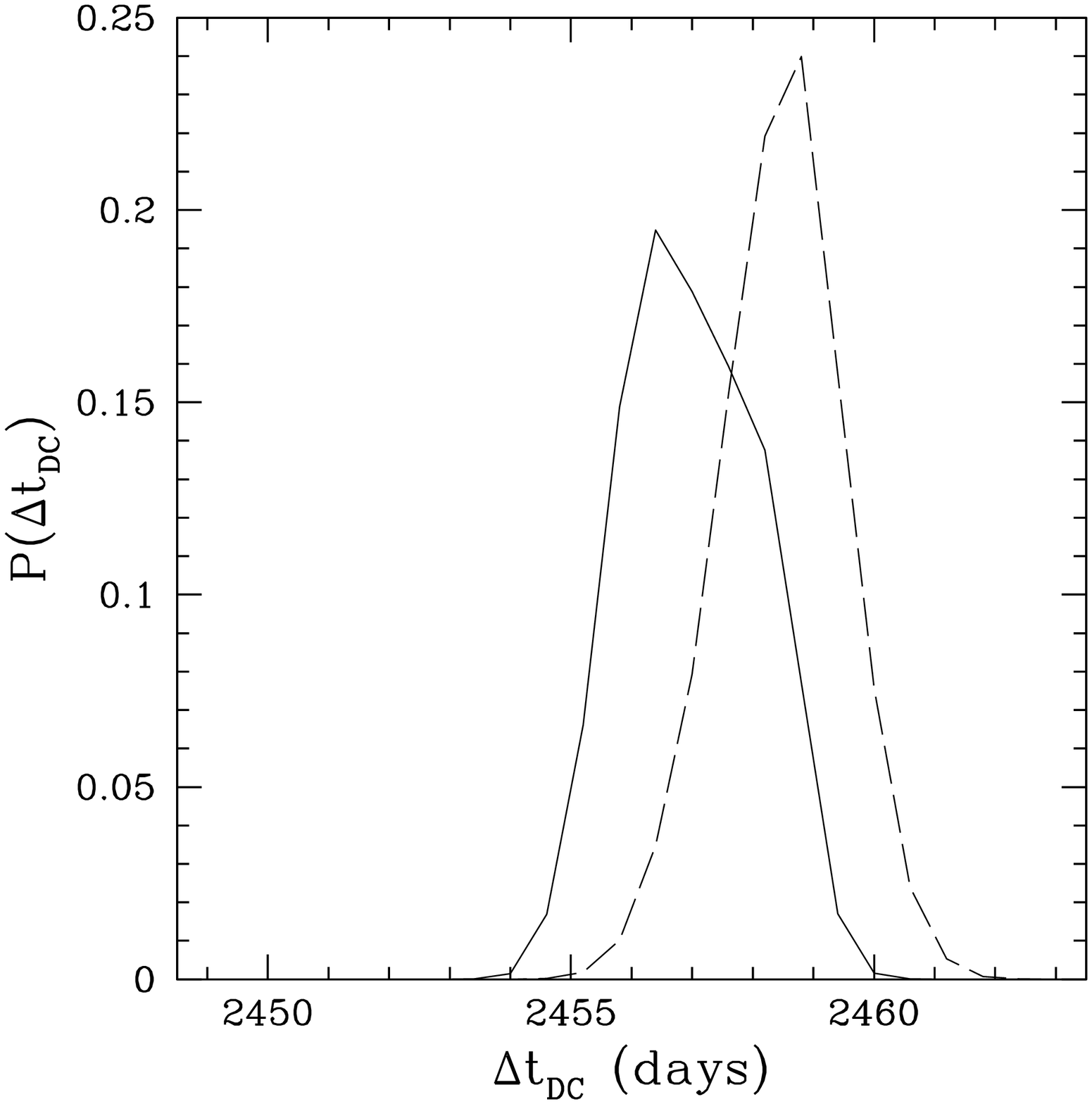}\\
\end{tabular}
\caption{\label{PDF} Probability distribution functions for the three independent time delays $\Delta t_{BC}$, $\Delta t_{AC}$ and $\Delta t_{DC}$ } built from the family of 75 models described in the text for the AIC (solid line) and BIC (dashed line) information criteria.
\end{figure}

The results for the 
four different ways of computing the lags are all in good agreement, albeit
not quite to the level of the reported statistical uncertainties.  For example,
the Joint BC, AC, DC, AB, DA and DB delays differ by $0.6$, $1.7$, $1.4$, $4.0$,
$2.1$ and $1.1\sigma$ using the average of the two statistical errors for $\sigma$.
The same is roughly true comparing the Joint and Pair results.  As seen in Fig.~2,
the probability distributions still substantially overlap. Thus, like essentially
all other time delay measurements, it would be best to use certainties several times the formal uncertainties to account for systematic uncertainties.  At least for the
absolute time delays, these uncertainties are irrelevant because of 
the cosmic variance in time delays due to fluctuations in the matter density along the line of sight.
For the models of \citet{1996ApJ...468...17B}, the expected cosmic variance is $\sim 4\%$ or  $31$, $33$, $98$,
$1.8$, $65$ and $57$~days for the BC through DB delays in Table~2. Even for the short AB delay, 
cosmic variance is the dominant uncertainty.

Delay ratios are far less affected by cosmic variance.  The cosmic variance is due to
the fluctuations in the surface density along the line of sight to particular lenses relative
to the mean background universe. However, a fluctuation $\kappa_{cv}$ in the convergence,  
which modifies the individual delays by $1-\kappa_{cv}$, has no effect on a delay ratio
because the effects on the two images cancel in a ratio.  While SDSS~J1004$+$4112 is a large separation lens,
it probably is not large enough for gradients in $\kappa_{cv}$ across the lens
to matter, so delay ratios are still limited by the statistical and
systematic errors in the measurements.  Because many of the delays are so long,
the fractional uncertainties in some of the delay ratios are tiny. For example
$\Delta t_{BC}/\Delta t_{DB} = 0.4668 \pm 0.0004$ - even if we double or triple
the formal uncertainty, the delay ratio is measured to $0.1\%$!

\begin{figure}
\begin{center}
\includegraphics[width=\textwidth]{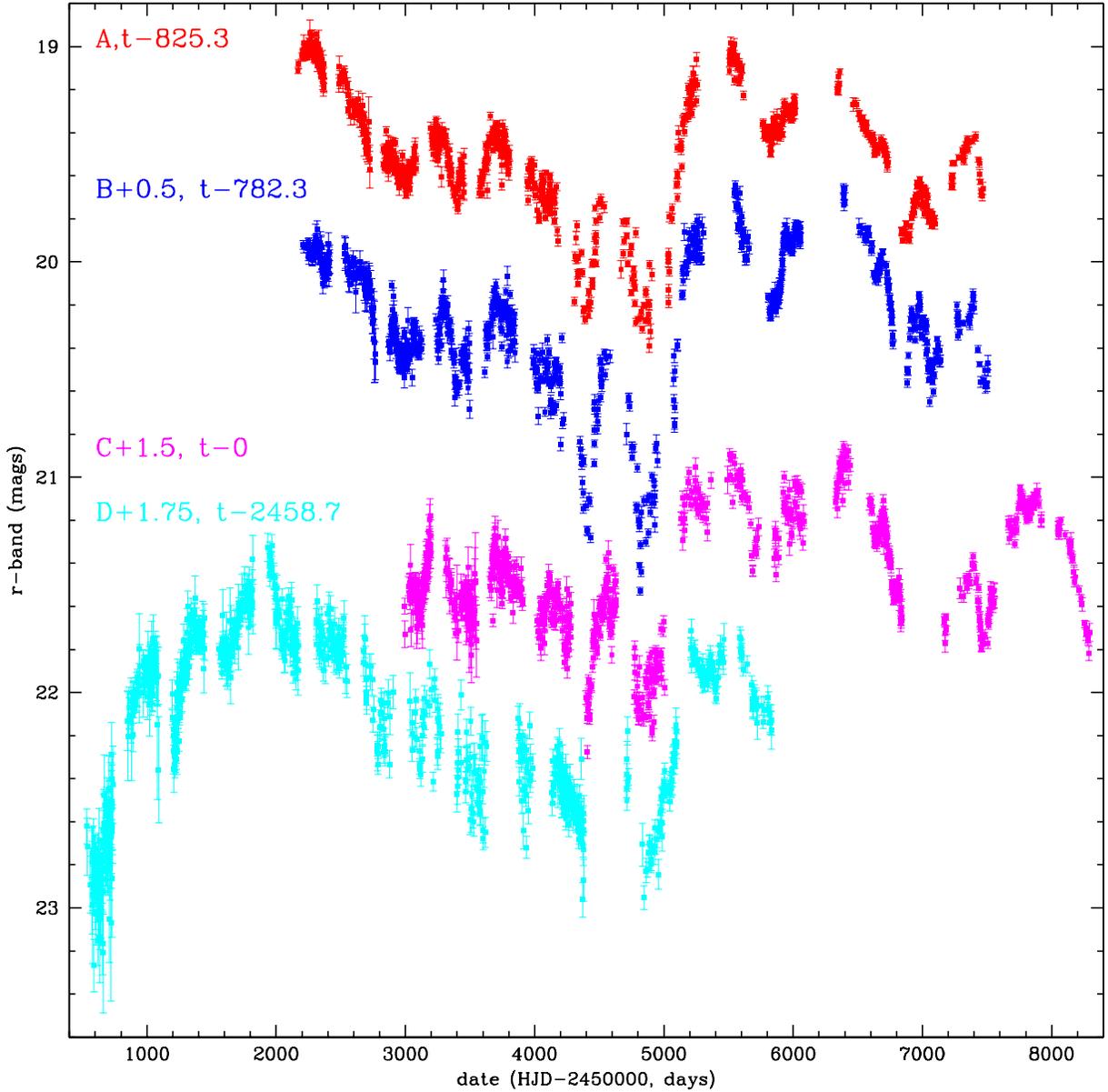}
\caption{\label{lcs} The  light curves from Fig. 
  1 shifted by time delays of 782.3 (B), 
  825.3 (A) and 2458.7 (D) days with respect to
  the leading image C.   We have added 0.5, 1.5 and 1.75
  magnitudes to the B, C and D images, respectively, for
  display purposes.}
\end{center}
\end{figure}

\begin{figure}
\begin{center}
\includegraphics[width=\textwidth]{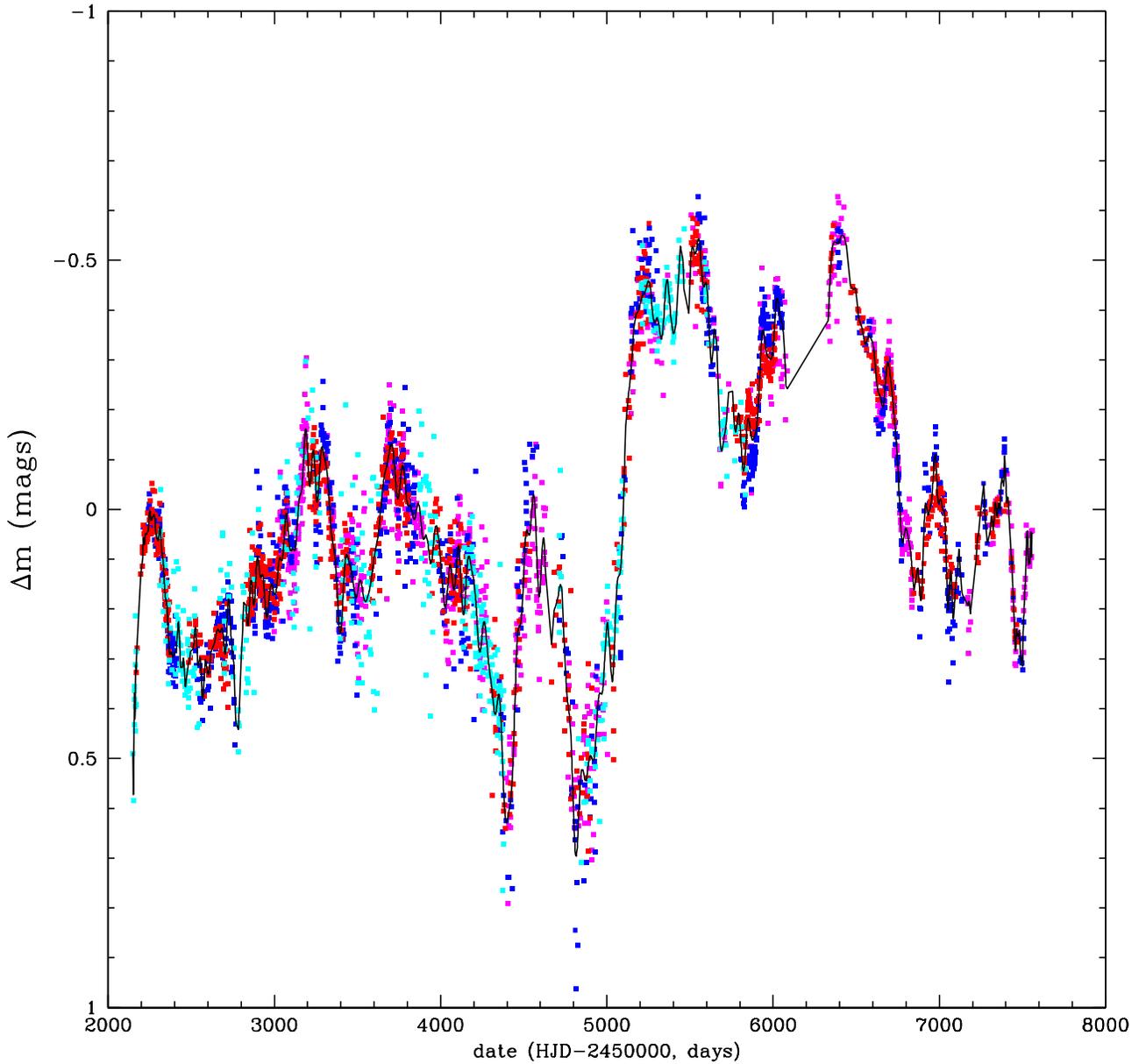}
\caption{\label{fits} The light curves shifted by the delays and with the microlensing
   polynomials subtracted.  The curve is the polynomial light curve for image C for
   the $n=350$, $m=15$
   model. The images have the same colors as in Figs.~1 and 3 (A red, B blue, C
magenta and D cyan). The photometric errors are omitted for clarity. }
\end{center}
\end{figure}

Fig.~3 shows the four light curves shifted by their $n=350$ and $m=15$ model delays relative to image C ($\Delta t_{BC}^{350.15}=782.3 \pm 0.3$ days,
$\Delta t_{AC}^{350.15}= 825.3 \pm 0.3$ days and 
$\Delta t_{DC}^{350.15}= 2458.7 \pm 0.6$ days).
One can see by eye that there are many large amplitude (compared to the errors) brightness fluctuations that are providing the time delay constraints.  In particular, all four light curves contain the sharp rise seen near 5000~days.  Fig.~4 shows the light curves shifted by the lags and with the microlensing polynomials subtracted to show how well they overlap.   This again emphasizes the large number of coherent variations sampled by multiple images as well as the way the large lags lead to a global light curve with no gaps over the time span where they overlap.

\begin{figure}
\begin{center}
\includegraphics[width=\textwidth]{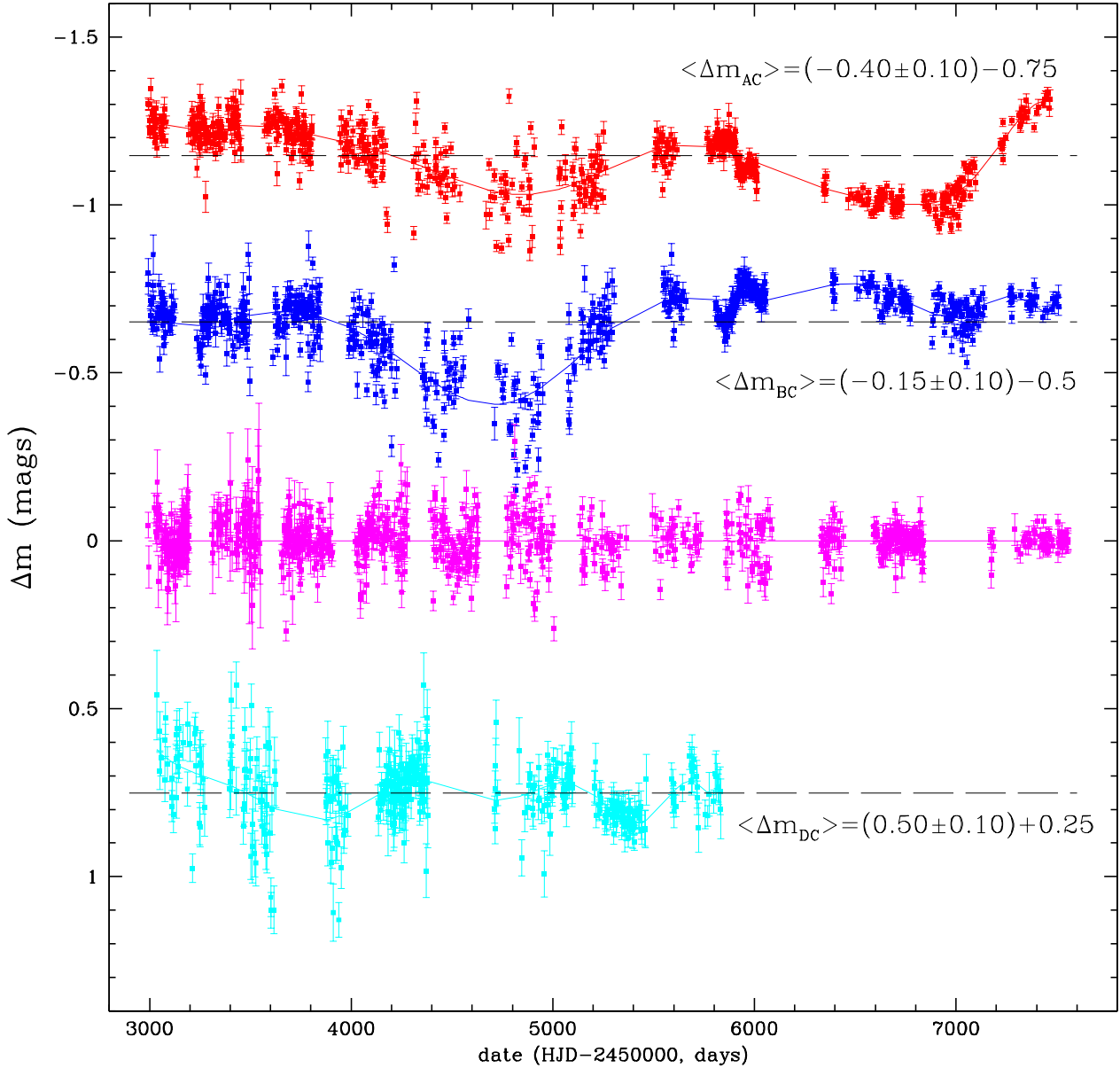}
\caption{\label{microlensing} The differential microlensing light curves after subtracting the $n=350$ polynomial model for the light curve of image C. For images A, B and D, the solid curves are the three $m=15$ order polynomials used to  model the differential microlensing between these image and C.  For image C, the mean residual relative to the light curve model is zero by definition.
The time
delays have been removed and we only show the period where the light curves overlap and the microlensing signal is well-defined. 
The images have the same colors as in Figs.~1, 3 and 4 (A red, B blue, C magenta and D cyan). The labels
$\langle \Delta m_{DC}\rangle = (0.50\pm0.10) +0.25$ give the mean magnitude offset, the dispersion of the residuals about the mean and then the shift in magnitude used to improve the visibility of each image.}
\end{center}
\end{figure}

Fig.~5 shows the differential microlensing of images A, B and D with respect to the reference 
image of C over the time period when they overlap.  The mean magnitude differences between 
image C and the time shifted images A, B and D are 
$-0.40 \pm 0.10$, $-0.15 \pm 0.10$ and 
$0.50 \pm 0.10$, respectively. The errors are the dispersions about the mean rather than
the uncertainties in the mean.  
The differences for A and B are very similar to those measured by \cite{2008ApJ...676..761F}.  
While there are some coherent short time scale features that
may be poorly modeled with intrinsic variability, there are clear, long time scale shifts 
in the image flux ratios that must be due to microlensing.  The little short time
scale structure, seen in the residuals, demonstrates that the assumption that we needed
a high order polynomial for the source variability but only a low order polynomial for
the microlensing model was justified.
The amplitude of the microlensing is modest, with a maximum amplitude of roughly
$0.5$~mag.  It is hard to impute the microlensing effects to particular images -- for
example, the similarity of the image A and B curves suggests that the microlensing is
dominated by image C but the dissimilarity of the image D curve argues against this hypothesis.

\section{Discussion and Conclusions}\label{sec4}

We have measured the last
time delay for the four bright images of SDSS~J1004+4112.  Image D lags image C
by 2457 days (6.73 years), the longest measured delay of any gravitational lens.  The ability to obtain light
curves without seasonal gaps and to flag a sharp variability feature in image C, carry out a
reverberation mapping campaign using images A and B and then fill in any missing data with image D 
makes SDSS~J1004+4112 an interesting prospect for such a monitoring campaign. Of the three predictions made
after the the measurement of the first two delays by \cite{2007ApJ...662...62F} and \cite{2008ApJ...676..761F},
the measured delay lies only inside the very broad prediction of 1500 to 2700 days by \cite{2015PASJ...67...21M}
and is much longer than predicted by \cite{2009MNRAS.397..341L} and \cite{2010PASJ...62.1017O}.  \citet{2010PASJ...62.1017O} notes that time delay of image D is correlated with the inner slope of the 
dark matter halo, so it is worthwhile revisiting models for the system. 

Time delays have been measured for several other cluster lenses.  \cite{2013ApJ...764..186F} measured
a $744\pm 10$~days delay for the two bright quasar images in SDSS~J1029$+$2623, the largest separation
(22\farcs5) quasar lens (\citealt{2006ApJ...653L..97I}). The only detailed model of this system used
this delay measurement (\citealt{2013MNRAS.429..482O}).  While SDSS~J1029$+$2623 has a third image
(\citealt{2008ApJ...676L...1O}) that could be used as a check of the model, it is faint and close to a 
brighter image, making it challenging to measure the additional delays.  
\cite{2015ApJ...813...67D} measured delays of $47.7\pm 6.0$~days and 
$722 \pm 24$~days 
between images
AB and CA of the six image quasar lens
SDSS~J2222$+$2745 (\citealt{2013ApJ...773..146D}).  
The models by \cite{2013ApJ...773..146D} had predicted an AB delay of $-87^{+187}_{-296}$~days and a CA delay of $1399^{+776}_{-850}$~days.
\cite{2017ApJ...835....5S} produced updated models
including the first two delay measurements, and it will be interesting to see how well they agree with future
measurements.  There are also two cluster lenses with lensed supernovae, with measured time delays for one. Predictions (\citealt{2016MNRAS.456..356D}, \citealt{2016MNRAS.457.2029J}, \citealt{2015MNRAS.449L..86O}, \citealt{2015ApJ...800L..26S},
\citealt{2009ApJ...703L.132Z}) for the time delay of supernova ``Refsdal'' (\citealt{2015Sci...347.1123K}) did encompass the eventually measured value (\citealt{2016ApJ...819L...8K}).
However, the predictions also spanned over 400 days and many of the models disagreed in their predictions.
It will be interesting to see how well the predictions for the decades long time delays of the second cluster lensed supernova, 
AT~2016jka (\citealt{2021NatAs...5.1118R}), hold up.

As in \cite{2007ApJ...662...62F} and \cite{2008ApJ...676..761F}, we again find that the light curves of the four images are not identical and thus that the images are being microlensed by stars associated with either the cluster galaxies or freely orbiting in the cluster.  The differential amplitudes relative to image C are a few tenths
of a magnitude, with slow variations over the $\sim 13$~years of overlap.
The microlensing was previously used by \cite{2008ApJ...676..761F} and \cite{2016ApJ...830..149F} to estimate the size of the quasar accretion disk.  The effects of microlensing are also seen in the broad emission line
profiles (\citealt{2004ApJ...610..679R}, \citealt{2006ApJ...645L...5G}, \citealt{2020A&A...634A..27P}) and the overall wavelength dependence of the quasar flux ratios (\citealt{2006A&A...454..493L}).  SDSS~J1004+4112 is interesting for microlensing because of the shorter microlensing time scales created by the high dynamical velocities of a cluster (\citealt{2011ApJ...738...96M}) and the prospect of observing microlensing from intracluster stars as opposed to stars associated with the cluster galaxies.

\section*{Acknowledgements}
The authors thank all of the astronomers who carried out the observations obtained
prior to the telescope shifting to robotic operations.
JAM is supported by the Spanish Ministerio de Ciencia e Innovaci\'on with the grant PID2020-118687GB-C32 and by the Generalitat Valenciana with the project of excellence Prometeo/2020/085.
CSK is supported by NSF grants AST-1908570 and AST-1814440.

\section*{Data Availability Statement}

The photometry used in the analysis is included in Table~1.

\clearpage


\begin{deluxetable}{cr@{ $\pm$ }lr@{ $\pm$ }lr@{ $\pm$ }lr@{ $\pm$ }l} 
\tablecaption{\label{tab:lc} Light curves in r-band magnitude for the four quasar images.}
\tablehead{JD-2450000 & \multicolumn2c{Image A} & \multicolumn2c{Image B} & \multicolumn2c{Image C} & \multicolumn2c{Image D} } 
\startdata 
2993.523 & 19.111 & 0.015 & 19.421 & 0.020 & 20.099 & 0.038 & 20.872 & 0.081 \\ 
2997.344 & 19.083 & 0.021 & 19.429 & 0.029 & 20.229 & 0.063 & 20.966 & 0.132 \\ 
3022.606 & 19.044 & 0.015 & 19.436 & 0.021 & 20.006 & 0.044 & 21.143 & 0.132 \\ 
3031.920 & 19.054 & 0.013 & 19.476 & 0.018 & 20.054 & 0.037 & 21.038 & 0.097 \\ 
3032.920 & 19.027 & 0.013 & 19.409 & 0.017 & 19.990 & 0.041 & 21.001 & 0.109 \\ 
3033.913 & 19.020 & 0.013 & 19.430 & 0.017 & 19.992 & 0.046 & 20.996 & 0.122 \\ 
3034.916 & 19.033 & 0.014 & 19.459 & 0.020 & 20.002 & 0.034 & 21.140 & 0.102 \\ 
3035.909 & 19.037 & 0.013 & 19.458 & 0.019 & 20.070 & 0.044 & 21.239 & 0.133 \\ 
3037.742 & 18.984 & 0.039 & 19.450 & 0.060 & 19.909 & 0.097 & 21.194 & 0.315 \\ 
3043.854 & 19.013 & 0.025 & 19.452 & 0.037 & 20.209 & 0.078 & 21.008 & 0.171 \\
\enddata 
\tablecomments{Table \ref{tab:lc} is published in its entirety in the machine-readable format. A portion is shown here for guidance regarding its form and content.}
\end{deluxetable} 

\begin{deluxetable}{lCCCCCC}
\tablecaption{\label{delays} Time Delays}
\tablewidth{0pt}
\tablehead{Model & \Delta t_{BC} & \Delta t_{AC} & \Delta t_{DC} & \Delta t_{AB} & \Delta t_{DA} & \Delta t_{DB}\\
 & \mbox{(days)} & \mbox{(days)}  & \mbox{(days)}  & \mbox{(days)}  & \mbox{(days)}  & \mbox{(days)}  }
 \startdata
 Joint AIC & 781.92 \pm 0.44 & 825.99 \pm 0.42 & 2456.99 \pm 1.11   & 44.04 \pm 0.23 & 1630.99 \pm 1.14 & 1675.06 \pm 1.14 \\ 
 Joint BIC  & 782.20 \pm 0.43 & 825.23 \pm 0.46 &  2458.47 \pm 1.02 & 43.01 \pm 0.27 & 1633.23 \pm 0.97 & 1676.26 \pm 0.97 \\
 Pair AIC  & 781.14 \pm 0.34 & 826.40 \pm 0.63 & 2456.62 \pm 1.15 & 41.36 \pm 0.17 & 1628.75 \pm 0.87  & 1675.02 \pm 2.03 \\
 Pair BIC & 780.00 \pm 0.38 & 827.34 \pm 0.62 & 2453.59 \pm 1.28 & 43.46 \pm 0.24 & 1636.41 \pm 2.34 & 1678.20 \pm 1.63 
 \enddata
\end{deluxetable}



\begin{thebibliography}{}

\bibitem[Alard \& Lupton(1998)]{Alard1998} Alard, C. \& Lupton, R.~H.\ 1998, \apj, 503, 325 

\bibitem[Alard(2000)]{Alard2000} Alard, C.\ 2000, \aaps, 144, 363. 

\bibitem[Bar-Kana(1996)]{1996ApJ...468...17B} Bar-Kana, R.\ 1996, \apj, 468, 17 

\bibitem[Dahle et al.(2013)]{2013ApJ...773..146D} Dahle, H., Gladders, M.~D., Sharon, K., et al.\ 2013, \apj, 773, 146

\bibitem[Dahle et al.(2015)]{2015ApJ...813...67D} Dahle, H., Gladders, M.~D., Sharon, K., et al.\ 2015, \apj, 813, 67

\bibitem[Diego et al.(2016)]{2016MNRAS.456..356D} Diego, J.~M., Broadhurst, T., Chen, C., et al.\ 2016, \mnras, 456, 356

\bibitem[Fian et al.(2016)]{2016ApJ...830..149F} Fian, C., Mediavilla, E., Hanslmeier, A., et al.\ 2016, \apj, 830, 149

\bibitem[Fohlmeister et al.(2007)]{2007ApJ...662...62F} Fohlmeister, J., Kochanek, C.~S., Falco, E.~E., et al.\ 2007, \apj, 662, 62

\bibitem[Fohlmeister et al.(2008)]{2008ApJ...676..761F} Fohlmeister, J., Kochanek, C.~S., Falco, E.~E., et al.\ 2008, \apj, 676, 761

\bibitem[Fohlmeister et al.(2013)]{2013ApJ...764..186F} Fohlmeister, J., Kochanek, C.~S., Falco, E.~E., et al.\ 2013, \apj, 764, 186

\bibitem[G{\'o}mez-{\'A}lvarez et al.(2006)]{2006ApJ...645L...5G} G{\'o}mez-{\'A}lvarez, P., Mediavilla, E., Mu{\~n}oz, J.~A., et al.\ 2006, \apjl, 645, L5

\bibitem[Inada et al.(2003)]{2003Natur.426..810I} Inada, N., Oguri, M., Pindor, B., et al.\ 2003, \nat, 426, 810

\bibitem[Inada et al.(2005)]{2005PASJ...57L...7I} Inada, N., Oguri, M., Keeton, C.~R., et al.\ 2005, \pasj, 57, L7

\bibitem[Inada et al.(2008)]{2008PASJ...60L..27I} Inada, N., Oguri, M., Falco, E.~E., et al.\ 2008, \pasj, 60, 27.

\bibitem[Inada et al.(2006)]{2006ApJ...653L..97I} Inada, N., Oguri, M., Morokuma, T., et al.\ 2006, \apjl, 653, L97. doi:10.1086/510671

\bibitem[Jackson(2011)]{2011ApJ...739L..28J} Jackson, N.\ 2011, \apjl, 739, L28

\bibitem[Jauzac et al.(2016)]{2016MNRAS.457.2029J} Jauzac, M., Richard, J., Limousin, M., et al.\ 2016, \mnras, 457, 2029

\bibitem[Kawano \& Oguri(2006)]{2006PASJ...58..271K} Kawano, Y. \& Oguri, M.\ 2006, \pasj, 58, 271

\bibitem[Kelly et al.(2015)]{2015Sci...347.1123K} Kelly, P.~L., Rodney, S.~A., Treu, T., et al.\ 2015, Science, 347, 1123

\bibitem[Kelly et al.(2016)]{2016ApJ...819L...8K} Kelly, P.~L., Rodney, S.~A., Treu, T., et al.\ 2016, \apjl, 819, L8

\bibitem[Lamer et al.(2006)]{2006A&A...454..493L} Lamer, G., Schwope, A., Wisotzki, L., et al.\ 2006, \aap, 454, 493

\bibitem[Liesenborgs et al.(2009)]{2009MNRAS.397..341L} Liesenborgs, J., de Rijcke, S., Dejonghe, H., et al.\ 2009, \mnras, 397, 341

\bibitem[McKean et al.(2021)]{2021MNRAS.505L..36M} McKean, J.~P., Luichies, R., Drabent, A., et al.\ 2021, \mnras, 505, L36

\bibitem[Mohammed et al.(2015)]{2015PASJ...67...21M} Mohammed, I., Saha, P., \& Liesenborgs, J.\ 2015, \pasj, 67, 21

\bibitem[Mosquera \& Kochanek(2011)]{2011ApJ...738...96M} Mosquera, A.~M. \& Kochanek, C.~S.\ 2011, \apj, 738, 96

\bibitem[Oguri et al.(2004)]{2004ApJ...605...78O} Oguri, M., Inada, N., Keeton, C.~R., et al.\ 2004, \apj, 605, 78

\bibitem[Oguri et al.(2008)]{2008ApJ...676L...1O} Oguri, M., Ofek, E.~O., Inada, N., et al.\ 2008, \apjl, 676, L1

\bibitem[Oguri(2010)]{2010PASJ...62.1017O} Oguri, M.\ 2010, \pasj, 62, 1017

\bibitem[Oguri et al.(2013)]{2013MNRAS.429..482O} Oguri, M., Schrabback, T., Jullo, E., et al.\ 2013, \mnras, 429, 482

\bibitem[Oguri(2015)]{2015MNRAS.449L..86O} Oguri, M.\ 2015, \mnras, 449, L86

\bibitem[Ota et al.(2006)]{2006ApJ...647..215O} Ota, N., Inada, N., Oguri, M., et al.\ 2006, \apj, 647, 215

\bibitem[Popovi{\'c} et al.(2020)]{2020A&A...634A..27P} Popovi{\'c}, L. {\v{C}}., Afanasiev, V.~L., Moiseev, A., et al.\ 2020, \aap, 634, A27

\bibitem[Sharon et al.(2005)]{2005ApJ...629L..73S} Sharon, K., Ofek, E.~O., Smith, G.~P., et al.\ 2005, \apjl, 629, L73

\bibitem[Sharon et al.(2017)]{2017ApJ...835....5S} Sharon, K., Bayliss, M.~B., Dahle, H., et al.\ 2017, \apj, 835, 5

\bibitem[Richards et al.(2004)]{2004ApJ...610..679R} Richards, G.~T., Keeton, C.~R., Pindor, B., et al.\ 2004, \apj, 610, 679

\bibitem[Rodney et al.(2021)]{2021NatAs...5.1118R} Rodney, S.~A., Brammer, G.~B., Pierel, J.~D.~R., et al.\ 2021, Nature Astronomy, 5, 1118

\bibitem[Ross et al.(2009)]{2009ApJ...702..472R} Ross, N.~R., Assef, R.~J., Kochanek, C.~S., et al.\ 2009, \apj, 702, 472

\bibitem[Sharon \& Johnson(2015)]{2015ApJ...800L..26S} Sharon, K. \& Johnson, T.~L.\ 2015, \apjl, 800, L26

\bibitem[Williams \& Saha(2004)]{2004AJ....128.2631W} Williams, L.~L.~R. \& Saha, P.\ 2004, \aj, 128, 2631

\bibitem[Zitrin \& Broadhurst(2009)]{2009ApJ...703L.132Z} Zitrin, A. \& Broadhurst, T.\ 2009, \apjl, 703, L132

\end{thebibliography}
\end{document}